%% file: Quantum-Enhanced_Sagnac_Sensing.tex
\documentclass{optica-article}

\journal{opticajournal} 

\articletype{Research Article}

\usepackage{lineno,subcaption,wrapfig}

\begin{document}

\title{Squeezing Enhanced Sagnac Sensing based on SU(1,1) Quantum Interference}

\author{Michal Natan, Saar Levin and Avi Pe'er\authormark{*}}

\address{Department of Physics and BINA Institute of Nanotechnology, Bar-Ilan University, Ramat-Gan 52900, Israel}

\email{\authormark{*}avi.peer@biu.ac.il} 

\begin{abstract*} 
We present a simple and robust design for a squeezing-enhanced Sagnac interferometer that employs the concept of SU(1,1) interference to significantly surpass the classical sensitivity limit (shot-noise limit - SNL) in rotational sensing. By strategically placing an optical parametric amplifier (OPA) inside the Sagnac loop, light is automatically squeezed in both forward and backward directions of the loop, which enhances the detectability of a small phase. For measuring the squeezed quadrature, we explore two approaches: Direct detection of the output intensity, which is simple, but requires a high-efficiency photo-detector; and parametric homodyne with an additional OPA, which accepts practical detectors with no efficiency limitation, but is technically more complex. Our analysis demonstrates super-classical sensitivity under most realistic conditions of loss and detector inefficiency, thereby leveraging the resources of squeezing and the principles of SU(1,1) interference, while maintaining compatibility with standard Sagnac configurations. 
\end{abstract*}

\section{Introduction}
\begin{wrapfigure}{r}{0.5\textwidth}
    \centering
    \includegraphics[width=0.48\textwidth]{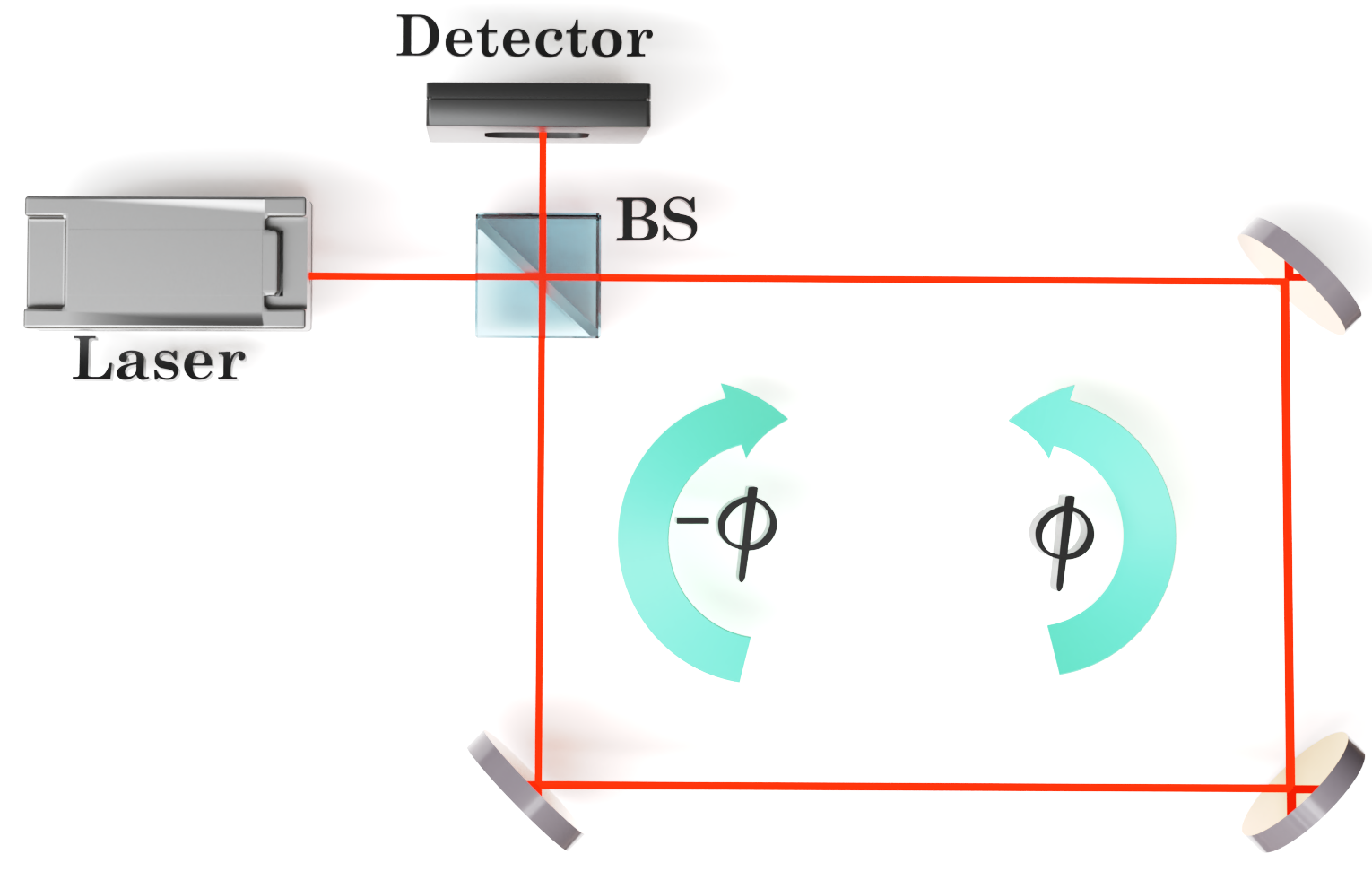}
    \caption{\textbf{The Sagnac Interferometer: Traditional Configuration.} A light beam is split and sent in opposite directions around the Sagnac loop. The two beams recombine to form a destructive interference pattern at the detector. When stationary, there's no phase shift between the beams, resulting in a consistent pattern. However, rotation causes a phase shift due to the differing path lengths traveled by the beams, altering the interference pattern. This shift is proportional to the rotation speed, making the Sagnac interferometer a precise tool for detecting rotational movements.}
    \label{fig:traditional_sagnac}
\end{wrapfigure}
The Sagnac interferometer is a major tool in optical technology, utilized primarily as an optical gyroscope for inertial navigation. Sagnac interferometers \cite{sagnac1913, post1967}, named after the French scientist Georges Sagnac, were first demonstrated for rotation sensing in 1913. Since then, they have been widely employed in various applications, such as inertial navigation \cite{shaked2018, moan2020}, north finding  \cite{prikhodko2013}, gravitational wave detection \cite{Eberle2010}, geophysical analysis \cite{stedman1997}, optical communication \cite{du2013}, lasing \cite{kuzin2001,ibarra2003} and sensing \cite{starodumov1997, ma2015}.

In a Sagnac interferometer, two light beams that are split from a single source travel through a ring in opposite directions along a common optical path, as depicted in Figure \ref{fig:traditional_sagnac}. Unlike other, well known interferometers, such as Mach-Zehnder or Michelson, the Sagnac design is inherently phase-stable Since the optical path is common to both arms, thereby automatically rejecting most phase fluctuations as common-mode and eliminating the need for phase control between different pathways. Consequently, Sagnac interferometers offer high stability against external disturbances like vibrations \cite{sagnac1913, post1967}, leading to simplified system complexity and enhanced stability.

The sensitivity of traditional interferometers (including the Sagnac) is limited by the quantum noise of the optical vacuum, also known as shot noise, which originates from the vacuum field injected through the unused input port of the beam splitter (BS). For a coherent state of light, the minimum detectable phase change is $\triangle\phi_{SNL} = \frac{1}{\sqrt{\langle \hat{N} \rangle}}$, dictated by the shot noise limit (SNL), where $\hat{N}$ is the number of photons that exit the interferometer during the measurement time. The primary objective of quantum interferometry is to harness quantum properties of the light, such as entanglement and squeezing \cite{bondurant1984, caves1985, schnabel2017, taguchi2020, bishop1988, lawrie2019, schonbeck2018} in order to surpass this classical limit, i.e. to measure minute optical phase shifts, below the SNL \cite{Eberle2010,caves1981, ou2012}.

Squeezed light, the major quantum resource for interferometric sensing, relies on a manipulation of quantum uncertainty: When two conjugate quantities (e.g. position and momentum in mechanics or the real and imaginary field-quadratures in optics) are relatively squeezed, the uncertainty of one component is reduced (squeezed), while the other is increased (stretched). Phase sensitive nonlinear processes in optics, such as parametric down-conversion and four waves mixing, can produce squeezed light. The utility of squeezing for quantum technology is diverse. For example, squeezing can improve the sensitivity of interferometric sensors by squeezing the uncertainty of the measured quantity, as utilized, for example by the LIGO and VIRGO detectors of gravitational waves \cite{grote2013, aasi2013, mcculler2020, miller2015, barsotti2018, acernese2019, acernese2020}.  

The common approach to harness squeezed light for improved sensing is by injection of squeezed vacuum into the unused port of a standard interferometer, such as Mach-Zehnder or Michelson \cite{schnabel2017, pezze2008} (commonly termed SU(2) interferometers, according to the symmetry group they adhere to), replacing the vacuum state and effectively reducing the vacuum quantum noise. This can lead to a significant decrease in total noise of the measurement below the shot noise threshold, thus improving the sensitivity of a phase measurement, which is proportional to the degree of squeezing \cite{wodkiewicz1985}. 

However, to harvest the reduction of noise in a realistic application requires extremely quiet detectors with near ideal quantum efficiency, along with electronic environments of ultra-low noise. This requirement is critical because any form of loss or inefficiency “contaminates” the light with additional vacuum noise, which is equivalent to a decoherence effect that significantly degrades the squeezing factor. Detectors of such high detection efficiencies at minimal electronic noise are technically demanding (and even unavailable in some wavelength ranges) \cite{schnabel2017}.

Alternatively, one can consider the SU(1,1) interferometer \cite{yurke19862, chekhova2016}, which marks a departure from traditional interferometric designs by substituting the linear BSs by OPAs. An OPA is a phase-dependent amplifier that amplifies one quadrature of the light, while squeezing the orthogonal quadrature, producing quadrature-squeezed quantum light, comprising two quantum-correlated spectral modes (known as ‘signal’ and ‘idler’), that have quantum correlations beyond the SNL in both intensity-difference $I_{1}-I_{2}$ and phase-sum $\phi_{1}+\phi_{2}$. The OPA can be thought of as a two mode squeezer that amplifies one quadrature of the combined signal-idler field and attenuates the other quadrature \cite{bello2021, schumaker1985}. 

Placing two OPAs in series (with a phase shift between them)  yields a phase-dependent output, similar to the interference output of a BS, but with key advantage over the SU(2) counterpart: the first OPA automatically generates the squeezed light with no need for an external squeezed seed, thereby yielding sub-shot-noise phase sensitivity \cite{Caves2020,Guo2018} (as long as the loss between the OPAs is sufficiently low). Additionally, an SU(1,1) interferometer is very robust to detection inefficiency / loss (because the light is amplified before the detection), allowing to operate with practical, inefficient detectors, and still enjoy the quantum advantage  
\cite{bello2021, chekhova2016, yurke19862, hudelist2014, lukens2017, li2016, chen2020}.  

\begin{figure}[h]
  \begin{subfigure}{.5\textwidth}
  \centering
    \includegraphics[width=1\linewidth]{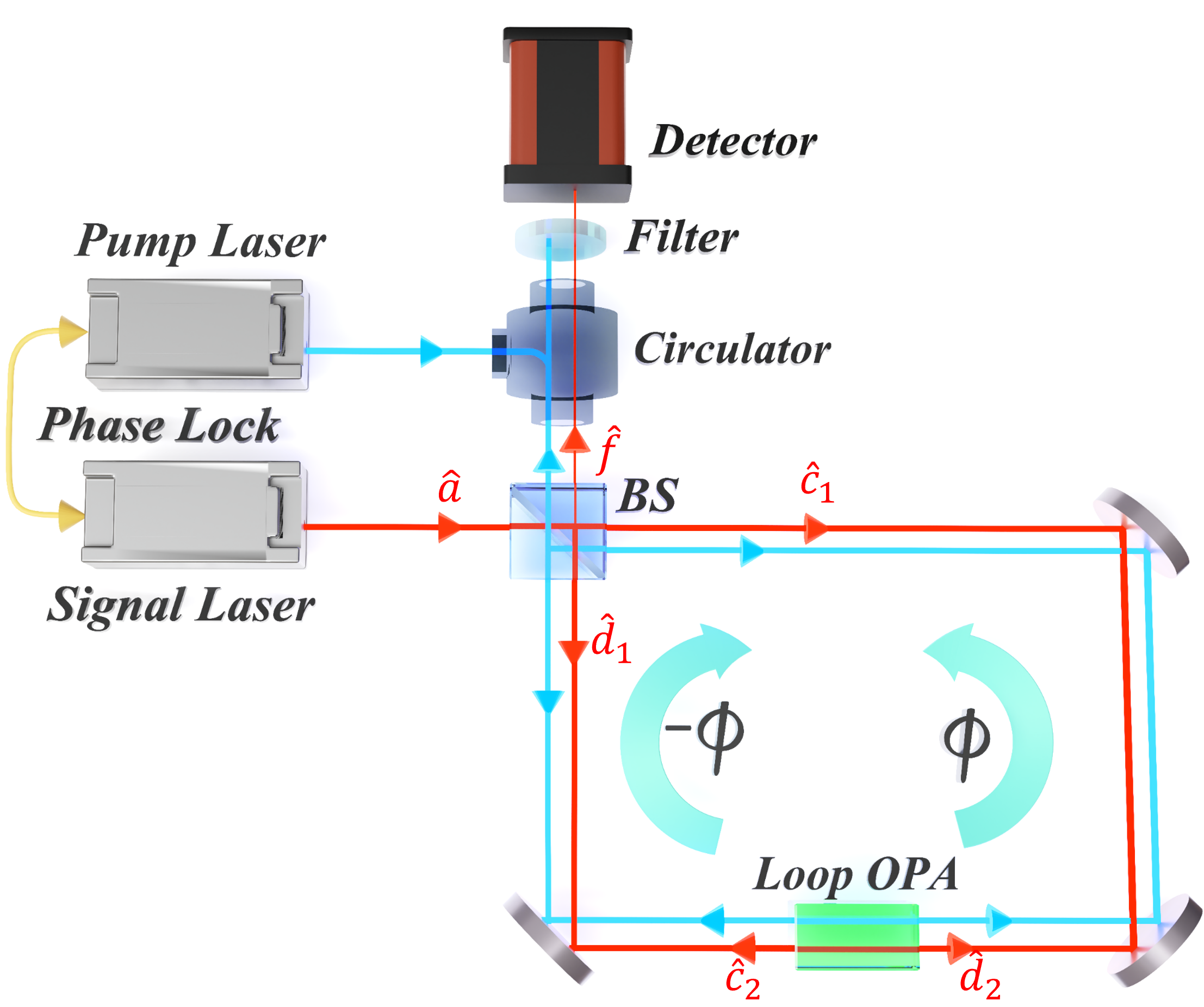}
    \caption{\textbf{Direct Intensity Detection.}}
  \end{subfigure}%
  \begin{subfigure}{.5\textwidth}
  \centering
    \includegraphics[width=1\linewidth]{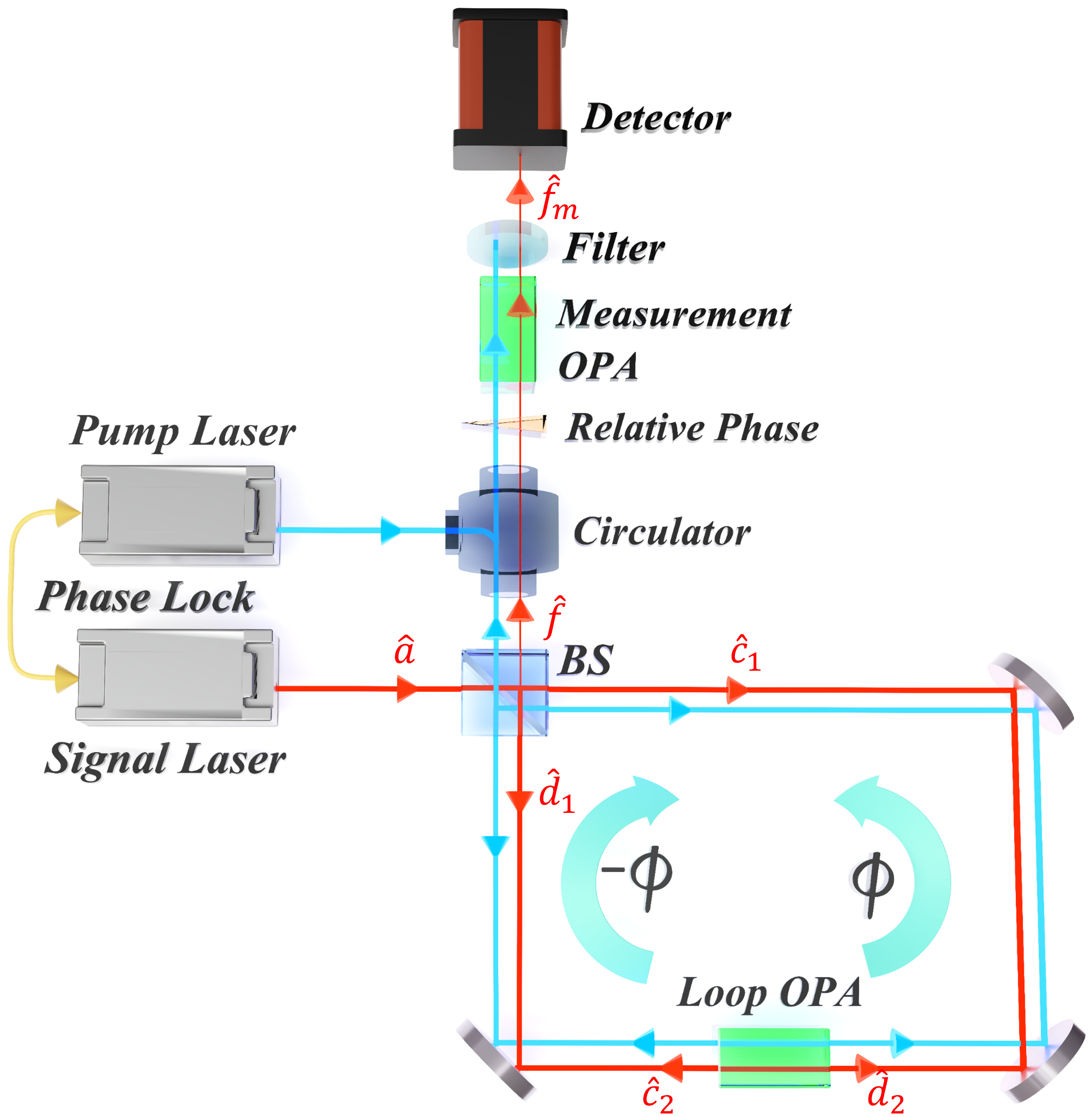}
    \caption{\textbf{Parametric Homodyne Detection.}}
  \end{subfigure}
  \caption{\textbf{Proposed squeezing-enhanced Sagnac sensor} with two alternative detection strategies:
\textbf{(a) Direct detection:} We incorporate into the standard classical Sagnac loop an OPA that squeezes the circulating fields in both directions. These squeezed beams experience a phase-difference in the loop (due to rotation) and recombine at the BS. Due to the interference between the two squeezed beams, the output at the dark port is intensity-squeezed (as we analyze and exemplify in  Fig. \ref{fig:quadrature_illustration}), allowing a direct readout of the rotation-induced phase information with sub-shot-noise sensitivity. This approach is technically simple, but requires an efficient photo-detectors, since detection losses directly degrade the observed squeezing. An advantage of this configuration is that the relative phase between the pump phase should be stabilized only at the input. If the pump for the loop OPA in both directions is combined through the Sagnac BS (either with the input seed or through the dark port), the squeezing phase is automatically maintained for both directions (as discussed in the text).  \textbf{(b) Parametric homodyne detection:} To alleviate the limitation of detector-efficiency, we add a second OPA at the dark output port for parametric homodyne detection. This measurement OPA is phased to amplify the squeezed quadrature of the output field, just before detection, which is highly robust against detector inefficiency, achieving super-classical sensitivity even with practical, lossy detectors.}

  \label{fig:modified_sagnac}
\end{figure}

The advantage of the SU(1,1) configuration in enhancing the phase sensitivity can be understood if we consider the minimum measurable phase shift  $\triangle\phi  = \frac{\triangle \hat{N}}{|\partial \hat{N} / \partial\phi|}$, where $\triangle \hat{N}$ is the noise of the detected intensity, and $\partial \hat{N}/ \partial\phi$ is the gradient of the measured intensity relative to the phase at the working point (phase) of the interferometer \cite{caves1981}. Thus, to improve the sensitivity, one must either reduce the measurement noise or increase the intensity gradient (or both). Here, the SU(1,1) approach holds a major advantage: while in an SU(2) interferometer (with a seed of squeezed vacuum) the output noise is reduced without altering the intensity gradient, SU(1,1) configurations amplify both the intensity noise and the intensity gradient, with the latter being amplified to a greater extent, thereby accommodating simpler, noisy, and lossy detectors. 

When turning to the Sagnac interferometer, one immediately identifies major differences compared to the standard SU(2) interferometers, such as Michelson or Mach Zehnder. Specifically, the Sagnac interferes two counter-propagating beams in the same loop. Here, unlike these traditional setups where a BS might be replaced with a parametric amplifier, such a direct substitution within a Sagnac interferometer proves ineffective \cite{Zha2023}, due to a distinctive characteristic of SU(1,1) interferometry, where the interference pattern is determined by the phase-sum of the two paths, rather than their phase difference. In the context of the Sagnac effect, rotation introduces to the two counter-rotating beams phase shifts of identical value but opposite sign ($\phi$ in one direction and $-\phi$ in the other, see Fig. \ref{fig:traditional_sagnac}). Thus, a simple replacement of the BS by an OPA will lead to a net cancellation of the rotational phase, making it insensitive to rotation. Consequently, a direct conversion of the Sagnac geometry into an SU(1,1) interferometer will not work.  Thus, previous attempts to harness  SU(1,1) interferometry for Sagnac sensing considered two OPAs in series within the Sagnac loop, forming a nested SU(1,1) interferometer, that required dual pumps with an independent loop, complex phase management, and a coherent homodyne detection stage at the output  \cite{Zha2023}. 

Here, we propose and analyze a simpler configuration with a single squeezing OPA within the Sagnac loop (see Fig. ~\ref{fig:modified_sagnac}), which simultaneously squeezes both counter-propagating beams. To detect the Sagnac phase with sub shot-noise sensitivity, we simply measure the light intensity at the dark output of the Sagnac (direct detection), as shown in Fig. \ref{fig:modified_sagnac}a. 
 
 A major advantage of this scheme is the pump evolution in the loop, which is automatically stabilized to the loop field. Because the pump and seed travel through the loop together, they experience the same phase shifts and their relative phase upon encountering the OPA is maintained as it was for the input (in both propagation directions), and the squeezing axis of the signal is automatically aligned. Thus, only a single external phase lock is needed to stabilize the seed-pump phase at the input. While this is clearly the simplest detection scheme possible, it requires near ideal photo-detectors with a quantum efficiency near unity, since the detection stage then becomes similar to standard homodyne, where losses in the detection stage introduce vacuum noise (shot-noise), and consequently, degrade the detection sensitivity.

To circumvent this requirement of high detection efficiency and allow operation with practical detectors, we suggest another measurement strategy using parametric homodyne detection: By incorporating an additional OPA at the output (dark) port of the Sagnac BS, we can amplify the optical quadrature of interest before it is detected (in the spirit of SU(1,1) detection \cite{caves1981,shaked2018,chekhova2016,bello2021,yurke19862,ou2012,Caves2020}). This approach, as depicted in Fig. \ref{fig:modified_sagnac}b, allows for precise extraction of the squeezed field-quadrature with standard, possibly inefficient photo-detectors, which is a significant practical advantage. 

Furthermore, phase management remains unaffected in the parametric homodyne configuration (in spite of the addition of the measurement OPA), since the pump travels along a common path with the seed, it will acquire opposite Sagnac phases ($\pm\phi$) that cancel upon recombination at the dark port (with no active phase lock required). The resulting pump phase remains stable for the measurement OPA, and while the intensity is modulated by $\cos^2{\phi}$, this effect is negligible for small Sagnac phases (our primary goal is to measure slow rotations), leaving the gain of the measurement OPA effectively constant.

Although this article analyzes mostly a degenerate Sagnac loop with a degenerate OPA (the signal frequency is exactly half the pump $\omega_s\!=\!\omega_p/2$) for the sake of simple and compact presentation, its principles apply fully also for non-degenerate, two-mode squeezing, where the signal and idler frequencies are distinct ($\omega_s+\omega_i=\omega_p$). Practically, two-mode operation enables simple, fiber-based configurations of the Sagnac loop, where the OPAs are implemented by four-waves mixing within the loop-fiber itself. In addition non-degenerate squeezing offers experimental flexibility of seeding only the signal/idler frequency, which removes completely the need to stabilize the phase/frequency of the Sagnac-seed and the pump for parametric-homodyne detection - a major practical simplification. A complete analysis of the non-degenerate configuration is provided in the supplementary material. 

The rest of this paper is organized as follows: Section ~\ref{sec:theoretical_evolution} develops the theoretical framework of the squeezing enhanced Sagnac interferometer, deriving the output operators from the input fields and introducing the two readout strategies: direct detection (Subsection ~\ref{sec:theoretical_dd}) and parametric homodyne detection (Subsection ~\ref{sec:theoretical_ph}).  Section ~\ref{sec:result} then presents our findings: Subsection ~\ref{sec:result_ph}  analyzes the phase sensitivity for parametric homodyne detection in a practical Sagnac sensor (with losses), starting with the ideal case of balanced gain and no losses, and extending to unbalanced gains and internal loop losses (both symmetric and asymmetric). Subsection ~\ref{sec:result_dd} examines direct detection on the same Sagnac configurations, focusing on the role of the seed and quantifying the impact of detection losses in this case. Finally, we briefly discuss the results for the non-degenerate case of two-mode squeezing, describing the differences from the degenerate case, and highlighting the opportunities it may hold for practical implementations in a fiber loop.

\section{Field evolution through the Sagnac loop}
 \label{sec:theoretical_evolution}
\begin{figure}[h]
    \includegraphics[width=1\linewidth]{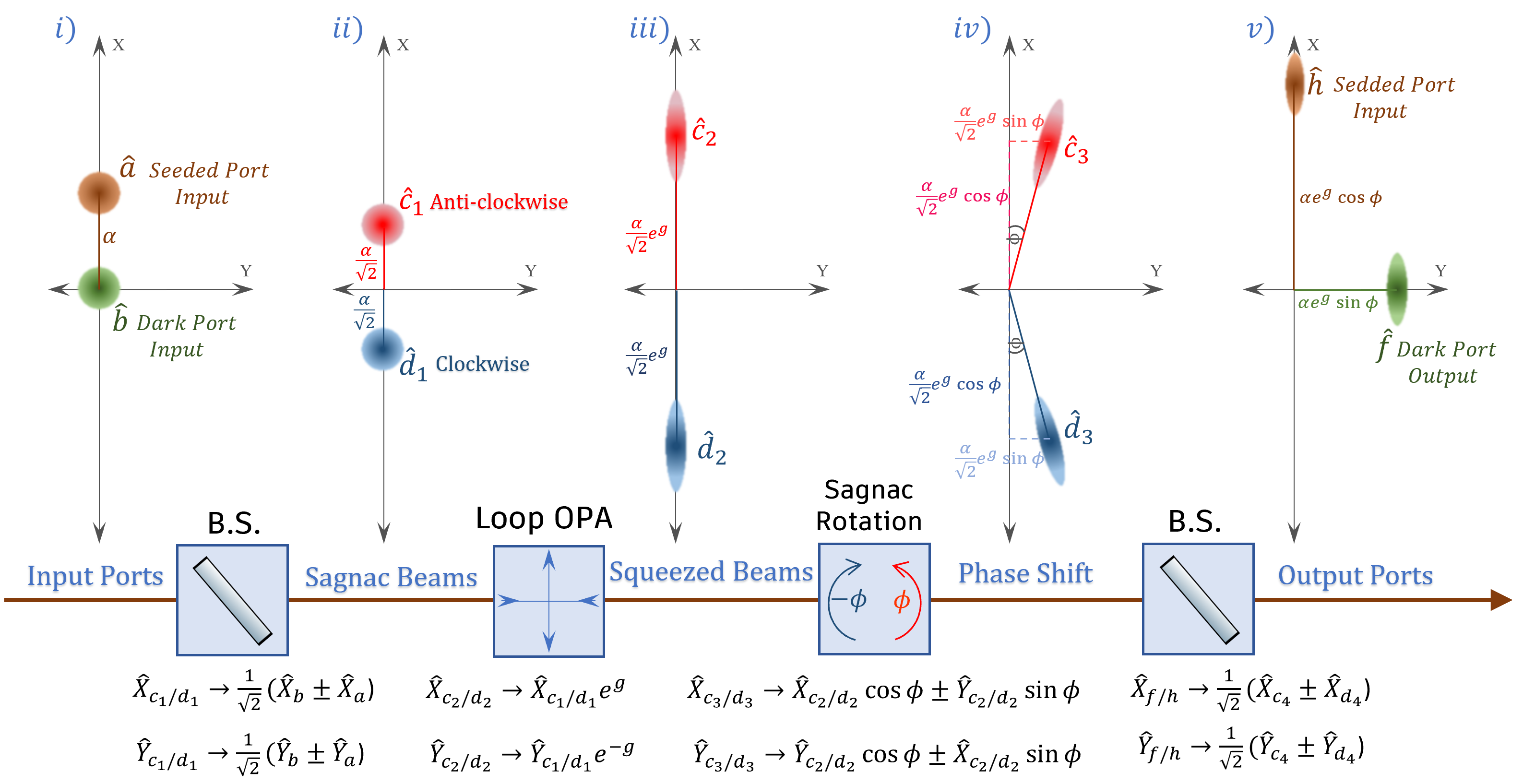}
        \caption{ \textbf{Quadrature evolution in the squeezing-enhanced degenerate Sagnac interferometer.}
  The figure illustrates the transformation of the optical quadratures $\hat{X}$ and $\hat{Y}$ at different stages of the interferometer.  
 \textbf{ (i)} \textbf{Input Ports:} A coherent seed with amplitude $\alpha$ enters the seeded port  $\hat{a}$, while vacuum $\lvert 0 \rangle$ enters the dark port  $\hat{b}$.  
 \textbf{ (ii)} \textbf{Sagnac Beams:} The input fields are split by the Sagnac BS into clockwise ($\hat{c}_{1}$) and counterclockwise ($\hat{d}_{1}$) propagating beams. The reflected beam acquires a $\pi$ phase shift relative to the transmitted beam (due to the input-output relations of a passive BS), leading to the destructive interference of the $\hat{X}$ quadrature at recombination.
  \textbf{(iii)} \textbf{Squeezed Beams:} Both beams  pass through the loop OPA, which squeezes the $\hat{Y}$-quadrature and amplifies the $\hat{X}$-quadrature. The quantum uncertainty distributions transform from circles into ellipses, reflecting the squeezing effect. 
 \textbf{ (iv)} \textbf{Phase Shift:} The counter-propagating beams accumulate opposite phase shifts $\pm \phi$ due to the Sagnac effect, yielding $\hat{c}_{3}$ and $\hat{d}_{3}$. (Propagation losses, not shown here for clarity, are modeled as effective BSs that couple the fields to vacuum modes, resulting in attenuated fields $\hat{c}_{4}$ and $\hat{d}_{4}$).  
  \textbf{(v)} \textbf{Output Ports:} Upon recombination at the BS, the output field at the dark port $\hat{f}$ is intensity-squeezed along the $\hat{Y}$ quadrature, whereas the bright output $\hat{h}$ (emitted back to the seed laser) is phase-squeezed along the $\hat{X}$ quadrature. Specifically, the $\hat{X}$ quadratures of fields $\hat{c}_{4}$ and $\hat{d}_{4}$ destructively interfere at the dark port, leaving the output $\hat{f}$ dominated by the squeezed $\hat{Y}$ quadrature, whose intensity carries the phase information (induced by the rotation).}
  \label{fig:quadrature_illustration}
\end{figure}

We now establish the theoretical framework for the modified Sagnac interferometer in the degenerate configuration, where the signal and idler fields share the same frequency.  This formalism applies to both detection methods: parametric homodyne and direct detection.  The quantum field operators at the various locations/stages along the interferometer are denoted  $a_x,b_x,c_x,d_x,f_x$, as marked in Fig. \ref{fig:modified_sagnac}, and then systematically propagate these fields through the various stages of the interferometer, as depicted in Fig. \ref{fig:quadrature_illustration}. By relating the output optical fields to the input fields, we can express any measurement operator in terms of the input and derive any measured quantity (e.g., average intensity and its fluctuations), which then allows us to calculate the sensitivity of the interferometer to small phase shifts.

A coherent seed beam is injected into the interferometer loop through one input port (the seeded port), initializing the field operator $\hat{a}$ to a coherent state $\lvert \alpha \rangle$. The second input port of the Sagnac BS, denoted by $\hat{b}$, admits a vacuum state $\lvert 0 \rangle$ (the dark port).  

The Sagnac BS ($R\!=\!T\!=\!50\%$) transforms the input fields into the clockwise ($\hat{c}_{1}$) and counterclockwise ($\hat{d}_{1}$) propagating fields:  
\begin{equation}
\begin{cases}
    \hat{c}_{1} = \dfrac{1}{\sqrt{2}}(\hat{a} + \hat{b}) \\[8pt]
    \hat{d}_{1} = \dfrac{1}{\sqrt{2}}(\hat{b} - \hat{a}).
\end{cases}
\end{equation}
The clockwise ($\hat{c}_{1}$) and counterclockwise ($\hat{d}_{1}$) fields propagate in opposite directions around the loop, both passing through the squeezing loop OPA. The transformation of these fields into $\hat{c}_{2}$ and $\hat{d}_{2}$  is governed by the input-output relation of the OPA (assuming a pump phase of zero): 
\begin{equation}
\begin{cases}
   \hat{c}_{2} = \hat{c}_{1}\cosh g + \hat{c}_{1}^{\dagger}\sinh g \\[8pt]
   \hat{d}_{2} = \hat{d}_{1}\cosh g + \hat{d}_{1}^{\dagger}\sinh g,
\end{cases}
\end{equation}
where $g$ is the parametric gain. After the OPA, the fields $\hat{c}_{2}$ and $\hat{d}_{2}$ continue through the interferometer in opposite directions along the common loop and acquire opposite phases due to the Sagnac effect, such that the field operators just before the Sagnac BS are: 
\begin{equation}
\begin{cases}
    \hat{c}_{3} = \hat{c}_{2} e^{i\phi} \\[8pt]
    \hat{d}_{3} = \hat{d}_{2} e^{-i\phi}.
\end{cases}
\end{equation}
Note that if the pump beam follows the Sagnac geometry of the seed (traveling along a common optical path), the phase relation between the fields in the loop OPA is automatically satisfied, requiring stabilization only at the Sagnac entrance.

To account for propagation losses, we introduce two additional BSs into the clockwise and counter-clockwise paths of the interferometer, modeling the coupling of each propagating field to an external vacuum mode. This approach captures both the attenuation of the optical signal and the injection of quantum noise from the environment, leading to: 
\begin{equation}
\begin{cases}
    \hat{c}_{4} = T_l\,\hat{c}_{3} + R_l\,\hat{l}_{1} \\[8pt]
    \hat{d}_{4} = T_l\,\hat{d}_{3} + R_l\,\hat{l}_{2},
\end{cases}
\end{equation}
where $\hat{l}_{1}$ and $\hat{l}_{2}$ are the vacuum noise operators associated with environmental coupling, and $T_l$ and $R_l$ are the transmission and reflection coefficients of the effective loss BSs, respectively, satisfying $T_l^2 + R_l^2 = 1$. We assume for simplicity that the losses after the loop OPA are symmetric for both beams in the loop (other options can be easily calculated as well). We do not need to consider the loss before the passage through the loop OPA, since the loop fields after the BS are classical, and therefore invariant to loss (coherent states remain coherent, only with a reduced amplitude, which can be calibrated into the input amplitude $\alpha$).

Finally, the fields $\hat{c}_{4}$ and $\hat{d}_{4}$ recombine at the same Sagnac BS, producing the dark-port output $\hat{f}$:
\begin{equation}
    \hat{f}= \dfrac{1}{\sqrt{2}}(\hat{d}_{4} + \hat{c}_{4}). 
\end{equation}
Let us clarify the total field-transformation through the Sagnac interferometer a bit further by explicitly writing the output quadratures of the dark port in terms of the input quadratures, For clarity, the following analysis considers the ideal, lossless configuration of the interferometer ($T_l=1, R_l=0$):
\begin{equation}
\begin{cases}
   \hat{X}_{f} =- \hat{X}_{b}e^{g}\cos{\phi} + \hat{Y}_{a}e^{-g} \sin{\phi}\\[8pt]
   \hat{Y}_{f} = \hat{Y}_{b}e^{-g}\cos{\phi} + \hat{X}_{a}e^{g}\sin{\phi}.
\end{cases}
\label{eq:quadrature}
\end{equation}
When the Sagnac phase is small ($\sin\phi\approx\phi, \cos^2\phi\rightarrow1, \sin^2\phi\rightarrow0$), and assuming a coherent seed at the seeded-port  with real amplitude $\alpha$ ($\hat{a} = \lvert \alpha \rangle \rightarrow \langle\hat{X}_a\rangle=\alpha$, $\langle\hat{Y}_a\rangle=0$) and vacuum at the dark-port($\hat{b} = \lvert 0 \rangle\rightarrow \langle \hat{X}_b \rangle\!=\!\langle \hat{Y}_b \rangle\!=\!0$),  the average quadrature amplitudes simplify to:
\begin{equation}
\begin{cases}
   \langle \hat{X}_f \rangle =0\\[8pt]
    \langle \hat{Y}_f \rangle = \phi\cdot\alpha e^{g},
\end{cases}
\end{equation}
showing the amplification of the Y quadrature that contains the phase information. 

At the same time, the variances of the quadratures are:
\begin{equation}
\begin{cases}
\Delta^2 \hat{X}_f = \tfrac{1}{4} e^{2g} \cos^2\phi 
          + \tfrac{1}{4} e^{-2g} \sin^2\phi \xrightarrow[\phi=0]{} \frac{1}{4}e^{2g} \\[8pt]
\Delta^2 \hat{Y}_f = \tfrac{1}{4} e^{-2g} \cos^2\phi 
          + \tfrac{1}{4} e^{2g} \sin^2\phi\xrightarrow[\phi=0]{} \frac{1}{4}e^{-2g}.
\end{cases}
\label{eq:var}
\end{equation}
which indicates that for small phases the $\hat{Y}$-quadrature is squeezed by $e^{-2g}$ below the SNL, while the $\hat{X}$-quadrature is stretched by $e^{2g}$.  Thus, the phase information encoded in the $\hat{Y}$ quadrature exhibits a quantum noise reduction relative to the SNL, enabling an enhanced sensitivity to phase. 

Intuitively, the interferometer acts as a natural quadrature filter: The two counter-propagating beams are squeezed by the loop OPA (Fig. \ref{fig:quadrature_illustration}iii) and then accumulate opposite Sagnac phase shifts, $+\phi$ and $-\phi$ (Fig. \ref{fig:quadrature_illustration}iv), which upon recombination at the BS (and taking into account the inherent $\pi$ phase shift of a passive, unitary BS), cause the components of the X-quadrature to interfere destructively at the dark output ((Fig. \ref{fig:quadrature_illustration}v). Thus, the classical amplitude at the dark-port is entirely at the Y-quadrature, which holds the information on the rotation-induced phase, with a squeezed noise, leading to the enhanced sensitivity beyond the SNL. From this point, the output field is detected according to the measurement method employed.

 \subsection{Direct Detection }
 \label{sec:theoretical_dd}
With direct detection, one simply performs an incoherent measurement of the intensity $\hat{f}^{\dagger}\hat{f}$ at the dark-port output of the interferometer without the need for a coherent homodyne detection or an additional OPA at the output. An intensity measurement can yield sub-shot noise sensitivity only if the detected light is intensity-squeezed. Luckily, this is indeed the case for the dark output due to the unique properties of the Sagnac interferometer with an internal OPA, as Eq. \eqref{eq:var} tells us and as illustrated in Fig. \ref{fig:quadrature_illustration}. 

From Eq. \eqref{eq:quadrature}, we can calculate the average intensity of the dark-port output $I_f= \langle\hat{f}^{\dagger}\hat{f}\rangle =\langle\hat{X}_{f}^{2}\rangle+\langle\hat{Y}_{f}^{2}\rangle-\dfrac{1}{4}$, using the quadrature intensities:
\begin{equation}
\begin{cases}
  \langle\hat{X}_{f}^{2}\rangle  
   = \dfrac{1}{4}(e^{2g}\cos^2{\phi} + e^{-2g}\sin^2{\phi}) 
  \;\rightarrow\; \dfrac{1}{4}e^{2g}, \\[8pt]
\langle\hat{Y}_{f}^{2}\rangle  
   = \dfrac{1}{4}e^{-2g}\cos^2{\phi} + (\dfrac{1}{4}+\alpha^2)e^{2g}\sin^2{\phi} 
   \;\rightarrow\;\alpha^2e^{2g}\phi^2,
\end{cases}
\end{equation}
where the right-arrow indicates the result for a sufficiently high seed intensity $|\alpha|^2\gg1$, moderately high gain $e^{2g}\gg e^{-2g}$, and a small, but non-zero phase $\phi\neq0$ (we will return later to the implication of this assumption), the dominant term of the intensity $\langle\hat{Y}_{f}^{2}\rangle =\alpha^2e^{2g}\phi^2\!$ reflects the Sagnac phase, but amplified by $e^{2g}$ relative to the classical case, whereas the other contributions to the intensity are dominated by amplified vacuum $\langle\hat{X}_{f}^{2}\rangle=\dfrac{1}{4}e^{2g}$.  Thus, for the total intensity of the dark output $I_f=\langle\hat{f}^{\dagger}\hat{f}\rangle =\langle\hat{X}_{f}^{2}\rangle+\langle\hat{Y}_{f}^{2}\rangle-\dfrac{1}{4}$ to faithfully represent the Sagnac phase, the condition $\langle\hat{Y}_{f}^{2}\rangle  \gg \langle\hat{X}_{f}^{2}\rangle $ must be satisfied, which translates to sufficiently large seed power.

Let us now show that quantum-enhanced sensitivity requires the power of the coherent seed ($\alpha^2$) to obey:
\begin{equation}
\alpha^{2} \gg \frac{e^{4g}}{8 \phi^2}.
\label{eq:seed_constrain}
\end{equation}
This condition becomes clear by analyzing the minimum detectable phase shift in the ideal, lossless configuration (shown in Fig. \ref{fig:Sensitivity_Direct_Detection}):
\begin{equation}
\Delta^2\phi_{DD} = \frac{e^{-4g}\sinh^2(2g)}{2\alpha^{4}}\left(\tan^{2}\phi + \cot^{2}\phi \right)
+ \frac{1}{\alpha^{2}} \left(e^{-4g} + \tan^{2}\phi \right).
\label{eq:sensitivity_DD}
\end{equation}
In the limit of moderately high gain, $e^{-4g}\sinh^2(2g) \rightarrow \tfrac{1}{4}$, and assuming a small phase ($\tan \phi \approx \phi$, $\cot \phi \approx \frac{1}{\phi}$), the phase sensitivity simplifies to
\begin{equation}
\Delta^2\phi_{DD} \approx \frac{e^{-4g}}{\alpha^{2}} +\frac{\phi^2}{\alpha^{2}}\left(1+\frac{1}{8\alpha^{2}}\right)+ \frac{1}{8\alpha^4\phi^2},
\end{equation}
which indicates that the minimum detectable Sagnac phase can be interpreted in terms of three physically distinct contributions that together determine the interferometer’s ultimate performance. The first term, proportional to $e^{-4g}/\alpha^{2}$, represents the squeezing-limited noise floor, governed by the reduced variance of the measured $\hat{Y}$ quadrature. The second contribution, scaling as $\phi^2$, corresponds to rotation noise, which increases at large Sagnac phases as the squeezing ellipse rotates away from the measurement basis. The third contribution, scaling as $\frac{1}{\phi^2}$, represents vacuum-noise contamination at very small Sagnac phases, where amplified vacuum fluctuations dominate the dark-port field. Luckily, this last term scales like $1/\alpha^4$, indicating that it is strongly suppressed at high seed power. 

To maintain genuine quantum enhancement, the squeezing-limited noise floor must dominate over the vacuum-noise contamination term, leading to the constraint of  Eq. ~\eqref{eq:seed_constrain} above $\alpha^{2} \gg \frac{e^{4g}}{8\phi^2}$.
This condition reveals that the required seed power grows exponentially with the parametric gain and becomes particularly demanding in the small-phase regime. Thus, a "sweet-spot" of minimal phase sensitivity is obtained for large enough seed at small Sagnac phases (though not exactly zero), as illustrated in Fig. \ref{fig:Sensitivity_Direct_Detection}, which shows the sensitivity enhancement relative to SNL. 

As mentioned, the SNL phase sensitivity of an interferometer with classical light is $\Delta^2\phi_{\mathrm{SNL}} = 1/{N_{\mathrm{in}}}$, where $N_{\mathrm{in}}$ is the total number of photons inside the interferometer, just before the output. For a fair comparison between the classical and the quantum case, we take $N_{in}$ to include also the classical amplification of the seed input by the OPA and the SPDC contribution to the intensity, resulting in:
\begin{equation}
\Delta^2\phi_{\mathrm{SNL}} = \frac{1}{N_{\mathrm{in}}} = \frac{1}{\alpha^{2} e^{2g}+2\sinh^{2}g }.
\label{eq:SNL}
\end{equation}
 Thus, the achievable quantum enhancement with direct detection relative to the shot-noise limit is:
\begin{equation}
\frac{\Delta^2\phi_{DD,min}}{\Delta^2\phi_{ \mathrm{SNL}}} 
= \frac{2\sinh^{2}g + \alpha^{2} e^{2g}}{\alpha^{2} e^{4g}}
\quad \xrightarrow[\alpha^{2}  \gg e^{4g}/8\phi^2]{} \quad e^{-2g}.
\end{equation}

Once the condition of Eq. \eqref{eq:seed_constrain} is met, the output intensity is dominated by the squeezed quadrature, and direct detection alone provides quantum-enhanced phase estimation. The primary advantage of this method is its simplicity, However, direct detection inherently requires a near-unit efficient photo-detector, which is a major limitation. Specifically, since the squeezed quadrature is only revealed at the detection stage, any detection inefficiency is interpreted as loss that directly injects vacuum noise and degrades the observable squeezing, thereby limiting the achievable phase sensitivity.

 \subsection{Parametric Homodyne Detection}
\label{sec:theoretical_ph}
To alleviate the requirements of both high-intensity seed and ideal, loss-free photo-detectors, we resort to parametric homodyne detection, where the dark-port output field $\hat{f}$ is further amplified before detection in a second OPA (measurement OPA), forming a nested SU(1,1) interferometer, such that $\hat{f}_{m} = \hat{f}\cosh g_{m}\! -\! \hat{f}^{\dagger}\sinh g_{m}$, with  $g_m$ the parametric gain and the pump phase chosen as $\phi_p\! =\! \pi$ so as to amplify the $\hat{Y}$-quadrature of interest. With sufficient gain $g_m\!\geq\! g$, the intensity of $\hat{f_m}^{\dagger}\hat{f_m}$ provides a direct readout of the amplified squeezed $\hat{Y}$-quadrature, which carries the Sagnac phase information.  

Crucially, the inherent phase stability of the Sagnac loop dramatically simplifies the operation of the measurement OPA: Because the pump travels the Sagnac loop with the signal/seed, the relative phase between the Sagnac output and the pump is passively stable in the measurement OPA. Thus, a simple passive phase-shifter will suffice to set the phase difference for the measurement OPA (using a dispersive element, like a wedge-prism),  obviating the need for active phase-stabilization.  Furthermore, even the phase stabilization for the loop OPA can be relaxed considerably, since phase fluctuations between the pump and the circulating degenerate field only reduce the amplification of the loop OPA slightly (generating slightly fewer photons), but the level of squeezing relative to shot-noise is maintained. Thus, as long as the pump–seed alignment is loosely maintained, the sensitivity enhancement relative to the SNL remains unchanged (with a slightly lower total flux of photons). Finally, we mention that the non-degenerate two-mode configuration enjoys here the ultimate simplification, as it requires \textit{no phase stabilization whatsoever} when only one of the input modes (signal or idler) is seeded (See analysis in the supplementary material). This is another major advantage for the non-degenerate configuration with parametric homodyne.  

In the ideal scenario of a lossless interferometer with balanced gain ($g_m = g$), the minimum detectable phase shift is now:
\begin{equation}
\Delta^2\phi_{PH}  = \frac{1}{\sinh^2(2g) + \alpha^2 e^{4g}}
+ \frac{\sinh^2(4g) + 2 \alpha^2 e^{8g}}{2\left[\sinh^2(2g) + \alpha^2 e^{4g}\right]^2}
\tan^2\phi .
\label{eq:sensitivity_PH}
\end{equation}

To evaluate the performance of this detection scheme, we compare it with the SNL sensitivity defined in Eq.~\eqref{eq:SNL}. The relative phase sensitivity is then:
\begin{equation}
\frac{\Delta^2\phi_{PH,min}}{\Delta^2\phi_{\mathrm{SNL}}} = 
\frac{2\sinh^{2}g + \alpha^{2} e^{2g}}{\sinh^{2}(2g) + \alpha^{2} e^{4g}} 
\;\approx_{\text{large } g}\;
e^{-2g}\,\frac{4\alpha^{2} + 2}{4\alpha^{2} + 1} .
\label{eq:sensitivity_min_PH}
\end{equation}
This comparison shows that the maximum phase sensitivity enhancement achievable with parametric homodyne detection at high coherent seeding is essentially the same as with direct detection: a factor of $e^{-2g}$ relative to the SNL. The key differences are that this enhancement can be achieved exactly at  $\phi=0$ and with much lower seed power. Even with no injected seed ($\alpha = 0$), the squeezed vacuum generated by the loop OPA is sufficient to enhance the sensitivity by  $\frac{1}{2\cosh^2{g}}=\frac{1}{2+\langle N_{in}\rangle}$, which shows Heisenberg-limited sensitivity \cite{Kolkiran07}.

The significant advantage of parametric homodyne detection is that the relevant quadrature is amplified before measurement, making the scheme inherently robust against detection inefficiencies. Losses occurring after the measurement OPA have little impact on the signal-to-noise ratio, enabling reliable sub-shot-noise phase sensitivity.

\section{Results}
 \label{sec:result}
\subsection{Parametric Homodyne Detection}
 \label{sec:result_ph}
We begin by analyzing the phase sensitivity of the squeezing-enhanced Sagnac interferometer compared to the SNL, under parametric homodyne detection. We start with the ideal, lossless scenario and continue to more realistic cases, accounting for propagation losses within the loop. 

\begin{figure}[h]
  \begin{subfigure}{.5\textwidth}
  \centering
    \includegraphics[width=1\linewidth]{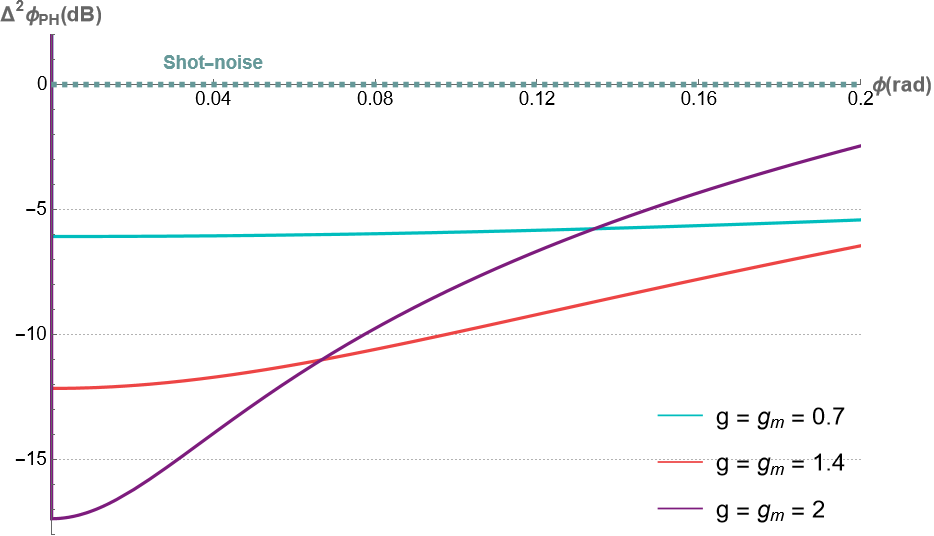}
    \caption{seeded configuration (lossless, $N_{seed}=|\alpha|^2=10$)}
  \end{subfigure}%
  \begin{subfigure}{.5\textwidth}
  \centering
    \includegraphics[width=1\linewidth]{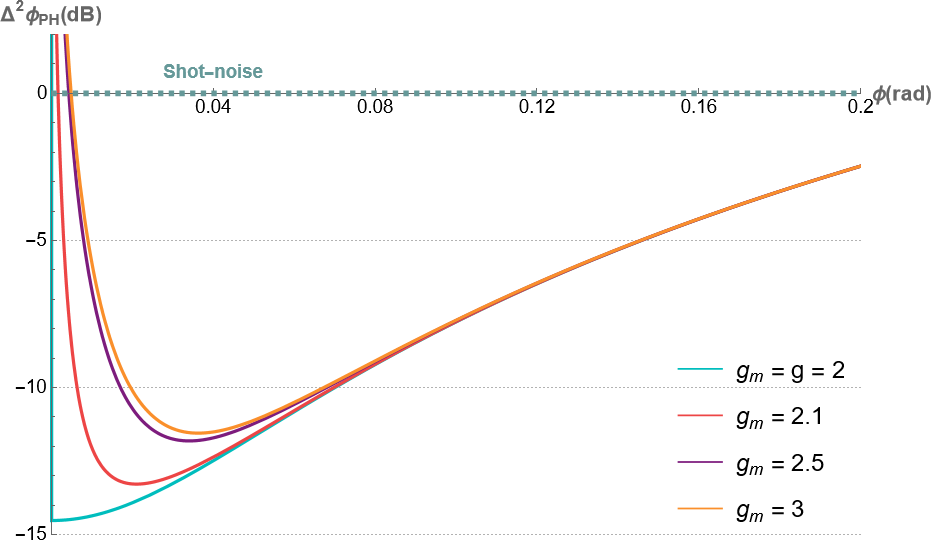}
    \caption{Unseeded configuration (lossless, $N_{seed}=0$)}
  \end{subfigure}
    \caption{Calculated phase sensitivity ($\Delta^2\phi_{PH}$) of our squeezing-enhanced Sagnac interferometer for an \textit{ideal, lossless configuration} with parametric homodyne detection. The Y- axis is log-scaled (dB) relative to the power-equivalent SNL, and the X-axis is the phase working-point of the interferometer. \textbf{(a) Seeded input}: We assume coherent seeding of $N_{seed}=10$ for different values of the parametric gain $g = g_m$ (balanced). \textbf{(b) Unseeded input}: ($N_{seed}=0$) with generation gain fixed at $g = 2$ and varying measurement gain $g_m$. The sensitivity is enhanced according to the generation gain, as expected. In the unseeded case (only), balanced gain is important, as imbalance degrades the enhancement and shifts the optimal working-phase away from zero}
 \label{fig:Sensitivity_balance_gain}
\end{figure}

In the lossless scenario with balanced gain ($g = g_{m}$), the interferometer consistently achieves sub-SNL phase sensitivity. As shown in Fig. ~\ref{fig:Sensitivity_balance_gain}a, the phase sensitivity follows Eq.~(\ref{eq:sensitivity_min_PH}), showing a quantum advantage that scales directly with the parametric gain $g$, as expected. Figure \ref{fig:Sensitivity_balance_gain}a shows the sensitivity assuming a coherent seed of $N_{seed}\!=\!|\alpha|^2\!=\!10$, which demonstrates optimal quantum-enhancement of $e^{-2g}$ relative to the SNL, already at such moderate seed levels.  The gain of the measurement OPA $g_m$ has little-to-no influence - once the measurement gain equals or exceeds the generation gain ($g_m \!\ge\! g$), the sensitivity curve simply does not change (showing $e^{2g}$ sensitivity enhancement near $\phi=0$) and indicating that the relevant quadrature has already been amplified sufficiently above the quantum noise floor.

Even when the seeding intensity is low (either few photons or completely unseeded $N_{seed} = 0$) the quantum enhancement persists, but is slightly reduced. Figure ~\ref{fig:Sensitivity_balance_gain}b shows the sensitivity for the unseeded (spontaneous) case, exhibiting enhancement that originates from the spontaneous parametric amplification of vacuum fluctuations. For balanced gain ($g_m = g$), the sensitivity near $\phi=0$ is enhanced by $\dfrac{1}{2\cosh^2{g}}=\dfrac{1}{\langle N_{in}\rangle+2}$, which demonstrates Heisenberg-scaling. Further increase of the measurement gain ($g_m > g$) unfortunately degrades this result in two aspects: The optimal operating point shifts away from $\phi = 0$ due to the additional noise from the measurement OPA at the dark fringe ($\phi=0$); and the sensitivity slightly degrades towards $4e^{-2g}$  for $g_m \approx 1.5g$  (increasing  $g_m$ further leaves the sensitivity unchanged).

\begin{figure}[h]
  \begin{subfigure}{.5\textwidth}
  \centering
    \includegraphics[width=1\linewidth]{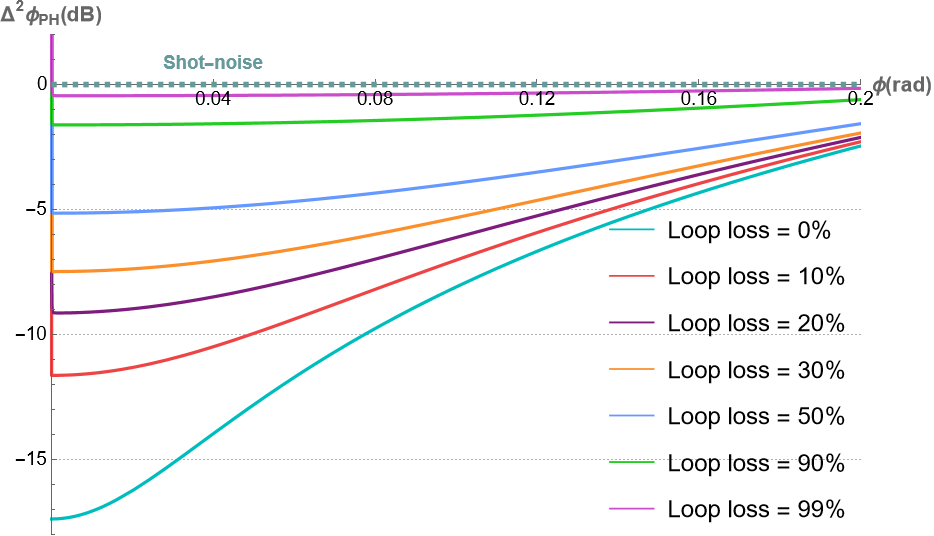}
    \caption{Symmetric loop loss}
  \end{subfigure}%
  \begin{subfigure}{.5\textwidth}
  \centering
    \includegraphics[width=1\linewidth]{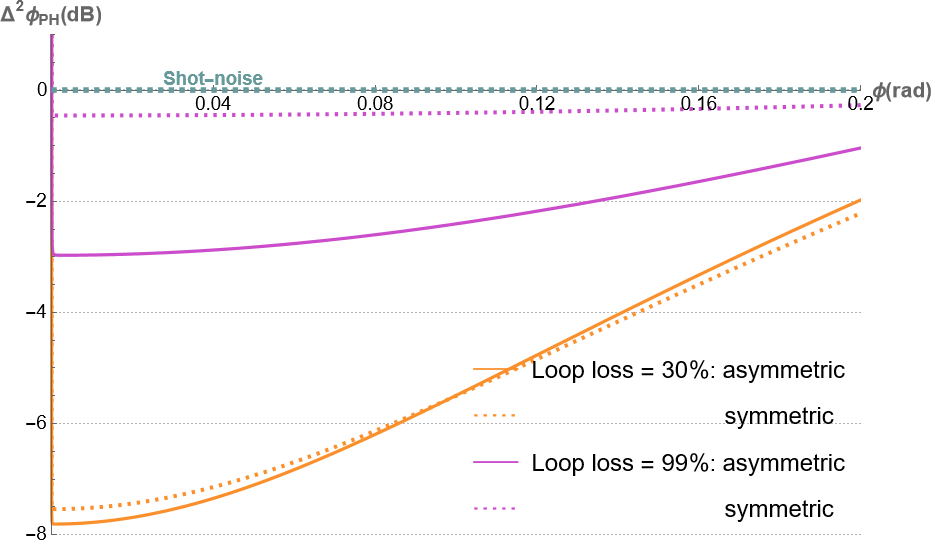}
    \caption{Symmetric vs asymmetric loop loss}
  \end{subfigure}
  \caption{\textbf{Effect of internal loss (symmetric \& asymmetric): }Calculated sensitivity  ($\Delta^2\phi_{PH}$ in dB relative to the SNL) with parametric homodyne detection, as a function of the phase working-point $\phi$. All graphs assume balanced gain $g=g_m=2$ and a coherent seed of $N_{seed}=100$  for two loss distributions within the Sagnac loop: \textbf{(a)} symmetric loss, where the loop loss before the loop OPA and after it are equal for both beams in the loop,  and \textbf{(b)} maximum asymmetric loss (solid line), where the loop OPA is placed near the BS, such that the forward (backwards) beam experiences all the loss before (after) the loop OPA. The dashed lines show the result with equivalent symmetric loss for comparison. For very large loop losses (we show 30\% and 99\% total loop loss) the maximally asymmetric loss is slightly better, but for reasonable (lower) loss situations the sensitivity becomes insensitive to the loss distribution.}
 \label{fig:Sensitivity_loop_losses}
\end{figure}

To analyze the performance of realistic practical Sagnac sensors (with reasonably high coherent seeding), we examine the effect of internal loss on the sensitivity enhancement in the balanced gain configuration ($g = g_{m}$). Propagation losses inside the Sagnac loop degrade the sensitivity by coupling vacuum fluctuations into the squeezed field, which "contaminates" the squeezed quadrature with noise. As shown in Figs.~\ref{fig:Sensitivity_loop_losses}a and ~\ref{fig:Sensitivity_loop_losses}b, increasing loss monotonically reduces the phase sensitivity enhancement and represents the major limiting factor for the achievable quantum advantage. For symmetric losses (both before and after the in-loop OPA), the phase-sensitivity enhancement relative to the SNL (near $\phi=0$ with sufficiently large gain) follows: 
\begin{equation}
\frac{\Delta^2\phi_{PH,\min}}{\Delta^2\phi_{\mathrm{SNL}}} 
\approx e^{-2g}T_l^{2}+R_l^{2},
\label{eq:sensitivity_deg_loss}
\end{equation}
where \(T_l\) and \(R_l\) denote the transmission and reflection coefficients of the internal loss channels, satisfying \(T_l^2 + R_l^2 = 1\). The total round-trip transmission of the loop is therefore  $T_l^4$, corresponding to an overall optical loop loss of \(1 - T_l^4\). While increasing loss gradually reduces the achievable enhancement, the interferometer maintains sensitivity that is equal to or better than the SNL even for extreme losses, up to 99\%.

Note that symmetric loss was assumed only for simplicity, and any other distribution of the loss along the loop can be analyzed just as well. Specifically, we compared symmetric loss (that is uniformly distributed in the loop before and after the loop OPA), and maximally asymmetric loss, where all loss is concentrated in one segment between the BS and the loop OPA. Both cases show very similar results, with a subtle distinction: In the symmetric configuration (Fig.~\ref{fig:Sensitivity_loop_losses}a), both arms are equally contaminated by loss before and after the loop OPA, so the dark-port interference combines two partially contaminated squeezed fields. In the maximally asymmetric configuration, one arm reaches the recombining BS as a nearly pure squeezed state, while the other is dominated by vacuum noise. As shown in Fig. ~\ref{fig:Sensitivity_loop_losses}b, this imbalance yields slightly better sensitivity at extreme losses: For example, at 99\% total losses, the asymmetric configuration outperforms the symmetric case by \~2.5 dB, but at moderate loss levels (e.g., 30\%), the difference between the two cases is very small (less than 0.4 dB). Thus, in practice, it is marginally favorable to position the squeezing OPA in a lossy loop as close as possible to the BS. 

\subsection{Direct Detection}
 \label{sec:result_dd}
In the direct detection configuration, the total intensity at the dark port, $\hat{f}^{\dagger}\hat{f}$, is measured directly, without any secondary amplifier, which is technically much simpler to implement. As analyzed in Sec.~\ref{sec:theoretical_dd} above, the performance of this single-OPA scheme can equal those of the corresponding parametric homodyne detection, under two additional conditions: First, an ideal, unit-efficiency photo-detector; and second, a sufficiently strong coherent seed of $\alpha^{2} \gg \frac{e^{4g}}{8\phi^2}$ (Eq. ~\ref{eq:seed_constrain}).
These conditions ensure that the phase-dependent signal in the squeezed $\hat{Y}$ quadrature dominates over the amplified vacuum fluctuations of the anti-squeezed $\hat{X}$ quadrature. Under this strong-seed condition, the phase sensitivity of direct detection matches that of parametric homodyne detection for two specific scenarios: the lossless case (as shown in Fig.~\ref{fig:Sensitivity_balance_gain}a) and the case involving internal propagation losses (as seen in Figs.~\ref{fig:Sensitivity_loop_losses}a and \ref{fig:Sensitivity_loop_losses}b). In both the lossless case and in the presence of symmetric internal propagation losses, the phase sensitivity follows the same scaling relation:
\begin{equation}
\frac{\Delta^2\phi_{DD,min}}{\Delta^2\phi_{SNL}} \approx e^{-2g}T_l^{2} + R_l^{2},
\end{equation}
reaching the maximum quantum enhancement of $e^{-2g}$ relative to the SNL in the lossless case.

\begin{figure}[h]
  \begin{subfigure}{.5\textwidth}
  \centering
    \includegraphics[width=1\linewidth]{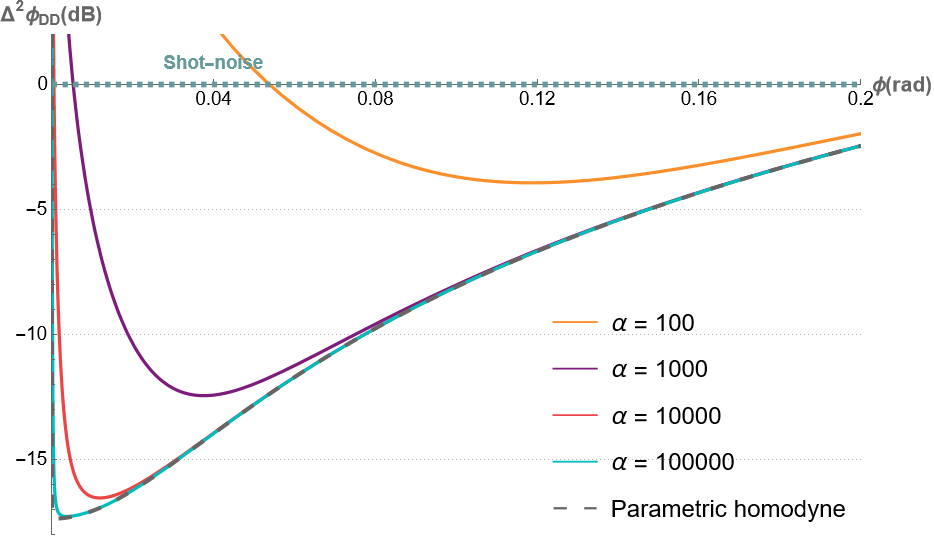}
    \caption{Impact of seeding-level}
  \end{subfigure}%
  \begin{subfigure}{.5\textwidth}
  \centering
    \includegraphics[width=1\linewidth]{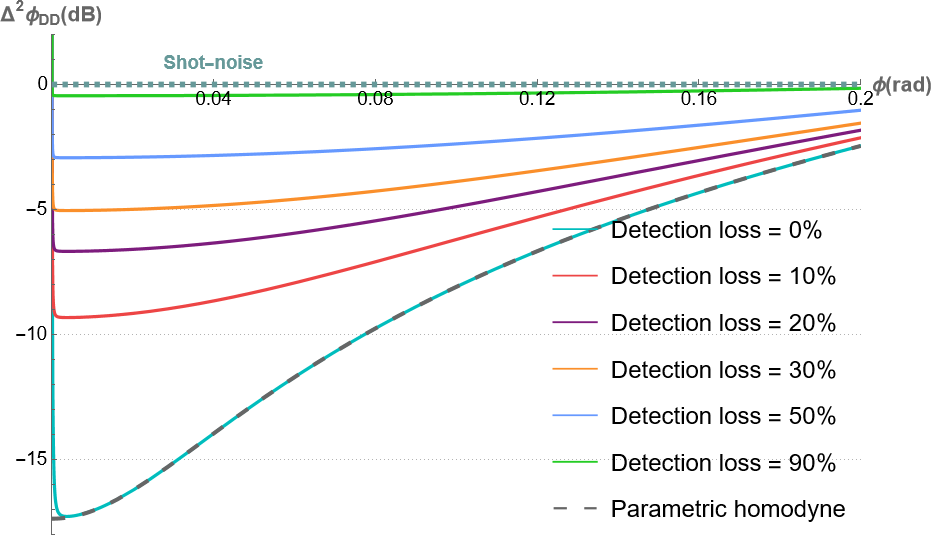}
    \caption{Impact of detection inefficiency (loss)}
  \end{subfigure} 
  \caption{Calculated sensitivity  ($\Delta^2\phi_{DD}$ in dB relative to the SNL) with direct detection, as a function of the phase working-point $\phi$. All graphs assume generation gain $g=2$ without internal loss.  For comparison, the dashed gray line marks the sensitivity of parametric homodyne detection.  \textbf{(a)} \textbf{Seed dependence} for ideal, unit-efficiency photo-detection. When the seed level ($N_{seed}=\alpha^2$) is increased, the sensitivity improves until saturation around $N_{seed}\sim 10^{10}$. \textbf{(b)} \textbf{Sensitivity for varying detection-loss} (with seed power $N_{seed} = 10^{10}$). Unlike parametric homodyne detection, direct detection is strongly affected by detection inefficiency.}
 \label{fig:Sensitivity_Direct_Detection}
\end{figure}

Fig.~\ref{fig:Sensitivity_Direct_Detection} shows the calculated performance of Sagnac sensing with direct detection: (a) shows dependence of the phase sensitivity  on the seed power  ($N_{seed}=\alpha^2$) (in the lossless case). For small seeds, the enhancement is negligible; as the seed power increases, the sensitivity rapidly improves and eventually saturates at $e^{-2g}$, which is the same quantum-enhanced level as the parametric homodyne scheme. This saturation reflects the transition to the regime where the dark-port intensity is fully dominated by the amplified $\hat{Y}$ quadrature; Fig.~\ref{fig:Sensitivity_Direct_Detection}b illustrates the effect of detection inefficiency: while the lossless case reaches the same enhancement as parametric homodyne detection, even moderate detection inefficiency (20–30\%) substantially reduces the quantum advantage, and at high detection losses the sensitivity approaches the SNL. With a practical detector of efficiency $T_d^2$ (detection loss $R_d^2=1-T_d^2$), the best phase-sensitivity enhancement relative to the SNL  $\frac{\Delta^2\phi_{DD,min}}{\Delta^2\phi_{SNL}} \approx T_d^2 e^{-2g} + R_d^2$, which is equivalent to that of symmetric internal losses.

\section{Conclusions}

We presented a simple and robust architecture for squeezing-enhanced Sagnac sensing based on SU(1,1) interference. The same theoretical framework naturally extends to the non-degenerate case, analyzed in detail in the Supplementary Material, which is particularly relevant for fiber-based implementations using four-wave mixing and offers a promising route toward fully integrated quantum-enhanced Sagnac sensors.

In the degenerate configuration, the common-path geometry ensures that both counter-propagating pump components share identical phase fluctuations, automatically preserving the correct pump phase inside the loop; the only remaining requirement is a loose external phase lock between the pump and the injected seed at the input. In the non-degenerate case, even this loos phase requirement can be lifted, as several practical configurations exist, where \emph{no phase locking is needed at all}. Specifically, when only one of the modes (signal or idler) is externally seeded, the partner mode is automatically generated inside the loop OPA  with the correct phase relative to the pump, enabling maximal photon-number amplification without active stabilization. Under these conditions, a parametric-homodyne–based configuration can reach the maximum SU(1,1) enhancement of $e^{-2g}$, just as in the degenerate case, and a direct-detection–based configuration can operate fully lock-free with only a 3dB penalty (reaching an enhancement of $2e^{-2g}$ relative to the SNL assuming a near-ideal detector).

Our analysis here considered a simple, single-pass Sagnac loop, where a high parametric gain was needed to produce considerable squeezing, This requirement can also be alleviated by configuring a multi-pass cavity around the Sagnac loop. We intend to analyze such cavity configurations in a future publication. 

\begin{backmatter}
\bmsection{Disclosures}
The authors declare no conflicts of interest
\end{backmatter}
\bibliography{references}

\clearpage
\section*{Supplementary Material}

\setcounter{section}{0}
\setcounter{equation}{0}
\setcounter{figure}{0}
\setcounter{table}{0}

\renewcommand{\thesection}{S\arabic{section}}
\renewcommand{\theequation}{S\arabic{equation}}
\renewcommand{\thefigure}{S\arabic{figure}}
\renewcommand{\thetable}{S\arabic{table}}

\input{supplement_content}

\end{document}

%% file: supplement_content.tex
\section{Field evolution through a non-degenerate Sagnac loop}
We now extend the theoretical analysis of the quantum Sagnac loop to include non-degenerate configurations, where the signal ($\omega_s$) and idler ($\omega_i$) modes are distinct and satisfy the energy conservation $\omega_s + \omega_i = \omega_p$. This configuration corresponds to two-mode squeezing and is experimentally relevant primarily for fiber-based Sagnac implementations, where optical parametric amplification can arise naturally from four-wave mixing within the loop fiber. This analysis applies to both detection schemes (direct / parametric), as discussed in the main text.

At the input port of the Sagnac interferometer, we have now two distinct fields $\hat{a}_{s,i}$, which can be either seeded coherently (with $\lvert \alpha_{s,i} \rangle$) or not, depending on the experimental choice. The other (dark) port of the Sagnac BS is fed by vacuum fields $\hat{b}_{s,i}$. After the 50:50 Sagnac BS, the clockwise ($\hat{c}_{s/i,1}$) and counterclockwise ($\hat{d}_{s/i,1}$) fields are:
\begin{equation}
\begin{cases}
    \hat{c}_{s/i,1} = \dfrac{1}{\sqrt{2}}(\hat{a}_{s/i} + \hat{b}_{s/i}) \\[8pt]
    \hat{d}_{s/i,1} = \dfrac{1}{\sqrt{2}}(\hat{b}_{s/i} - \hat{a}_{s/i}).
\end{cases}
\end{equation}
Both beams propagate in opposite directions around the Sagnac loop and interact with the loop OPA:
\begin{equation}
\begin{cases}
   \hat{c}_{s/i,2} = \hat{c}_{s/i,1} \cosh{g}+ \hat{c}_{i/s,1} ^{\dagger}\sinh{g}
   \\[8pt]
    \hat{d}_{s/i,2} =\hat{d}_{s/i,1} \cosh{g} + \hat{d}_{i/s,1} ^{\dagger} \sinh{g}, 
\end{cases}
\end{equation}
where $g$ is the gain of the loop OPA (assumed real positive). Due to the rotation, the counter-propagating fields acquire opposite Sagnac delays $\pm\tau$  that correspond to Sagnac phase shifts $\phi_{s,i}=\omega_{s,i}\tau$. Since the parametric amplification is governed by the pair-phase $\phi_s\!+\!\phi_i\!=\!\left(\omega_s\!+\!\omega_i\right)\tau\!=\!\omega_{p}\tau$, we simply assign $\phi\!=\!\frac{\phi_s\!+\!\phi_i}{2}\!=\!\frac{\omega_{p}\tau}{2}$ to both of them:
\begin{equation}
\begin{cases}
    \hat{c}_{s/i,3} = \hat{c}_{s/i,2} e^{i\phi}, \\[6pt]
    \hat{d}_{s/i,3} = \hat{d}_{s/i,2} e^{-i\phi}.
\end{cases}
\end{equation}
To account for propagation losses, we include effective BSs in both arms, coupling each propagating mode to vacuum noise fields $\hat{l}_{s/i,1}$ and $\hat{l}_{s/i,2}$:
\begin{equation}
\begin{cases}
    \hat{c}_{s/i,4} = T_l\,\hat{c}_{s/i,3} + R_l\,\hat{l}_{s/i,1} \\[6pt]
    \hat{d}_{s/i,4} = T_l\,\hat{d}_{s/i,3} + R_l\,\hat{l}_{s/i,2},
\end{cases}
\end{equation}
where $T_l^2 + R_l^2 = 1$. For simplicity, we assume that the losses are identical for both the signal and idler modes, and that the losses are symmetrically distributed in both propagation directions of the Sagnac loop (other options can be easily calculated as well). Finally, the two counter-propagating modes recombine at the Sagnac BS again, producing the dark-port outputs:
\begin{equation}
\begin{cases}
    \hat{f}_{s/i} = \dfrac{1}{\sqrt{2}}(\hat{d}_{s/i,4} + \hat{c}_{s/i,4}) \\[8pt]
    \hat{h}_{s/i} = \dfrac{1}{\sqrt{2}}(\hat{d}_{s/i,4} - \hat{c}_{s/i,4}).
\end{cases}
\end{equation}

The distinct signal and idler fields in this two-mode extension of the degenerate model offers greater experimental flexibility, e.g. in coherently seeding, or in measuring just one of them.

\subsection{Direct detection}

\begin{wrapfigure}{r}{0.5\textwidth}
    \centering
    \vspace{-8pt}
    \includegraphics[width=0.49\textwidth]{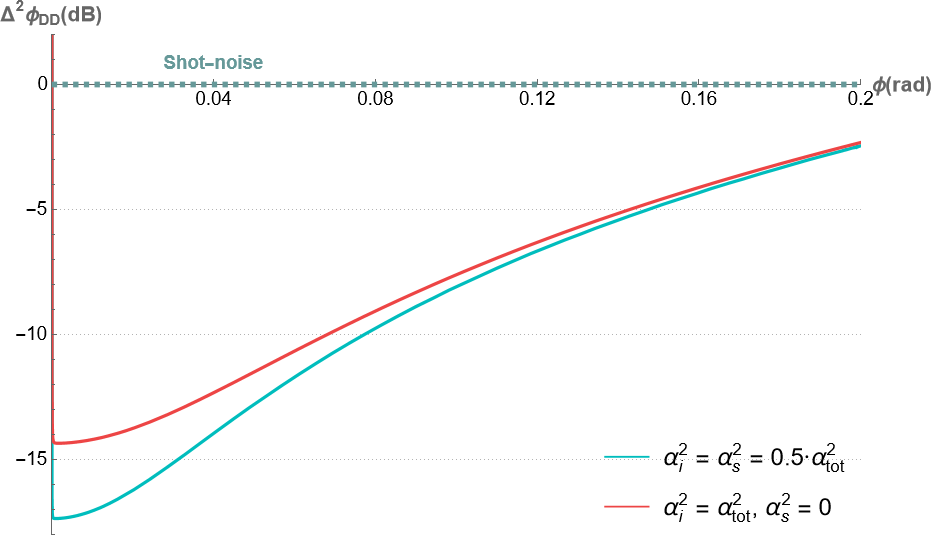}
    \caption{\textbf{Effect of seed distribution:} calculated phase sensitivity ($\Delta^2\phi_{DD,i+s}$ in dB relative to the SNL) of our non-degenerate squeezing-enhanced Sagnac interferometer for an ideal, lossless configuration with\textit{ direct detection}, as a function of the phase working-point $\phi$. Both signal and idler modes  $\langle \hat{f}_s^\dagger \hat{f}_s + \hat{f}_i^\dagger \hat{f}_i \rangle$ are measured for generation gain $g=2$ and large seed intensity $N_{seed} = 10^{10}$. With single-mode seeding ($\alpha_i=\sqrt{N_{seed}}$,  $\alpha_s=0$), the enhancement is limited to \(2e^{-2g}\) (red line). To obtain the maximum achievable enhancement of $e^{-2g}$ (green line),  the seed should be symmetric ($\alpha_i = \alpha_s$), which corresponds to a classical pure two-mode field that seeds only one two-mode quadrature (leaving the other quadrature at vacuum).}
    \label{fig:2_modes_direct_detection}
\end{wrapfigure}

In the non-degenerate Direct Detection scheme, the phase-sensitive information is extracted by jointly measuring the total intensity at the dark port of both the signal and idler $I_f = \langle \hat{f}_s^\dagger \hat{f}_s + \hat{f}_i^\dagger \hat{f}_i \rangle$. In calculating the phase sensitivity we followed the same procedure as in the degenerate-case, keeping signal–idler co-variance. 

The maximal quantum enhancement of the sensitivity is achieved only when the total seed power ($\alpha_{\mathrm{tot}}^2 = \alpha_s^2 + \alpha_i^2$) satisfies $\alpha_{\mathrm{tot}}^2 \gg \frac{e^{4g}}{8\phi^2}$. indicating that the phase-dependent contribution in the squeezed quadrature dominates the amplified vacuum, analogous to the degenerate case. Crucially, the distribution of the seed between signal and idler determines the maximal enhancement. If all the seed is injected into a single mode (signal or idler), the enhancement is limited to $2e^{-2g}$ (14.4\,dB for $g=2$). With the same total seed power but symmetric seeding ($\alpha_s=\alpha_i$),  the full quantum limit $e^{-2g}$ (17.4\,dB for \(g=2\)) is achieved, as illustrated in Fig.~\ref{fig:2_modes_direct_detection}. , corresponding to a two-mode seed of a classical pure two-mode quadrature (leaving the other quadrature at vacuum). Thus, while the maximum enhancement is the same as in the degenerate case, the non-degenerate scheme requires a specific $\alpha_s = \alpha_i$ condition to fully realize the $e^{-2g}$ benefit.

\subsection{Parametric homodyne detection}

\begin{figure}[h]
    \begin{subfigure}{.5\textwidth}
        \centering
        \includegraphics[width=1\linewidth]{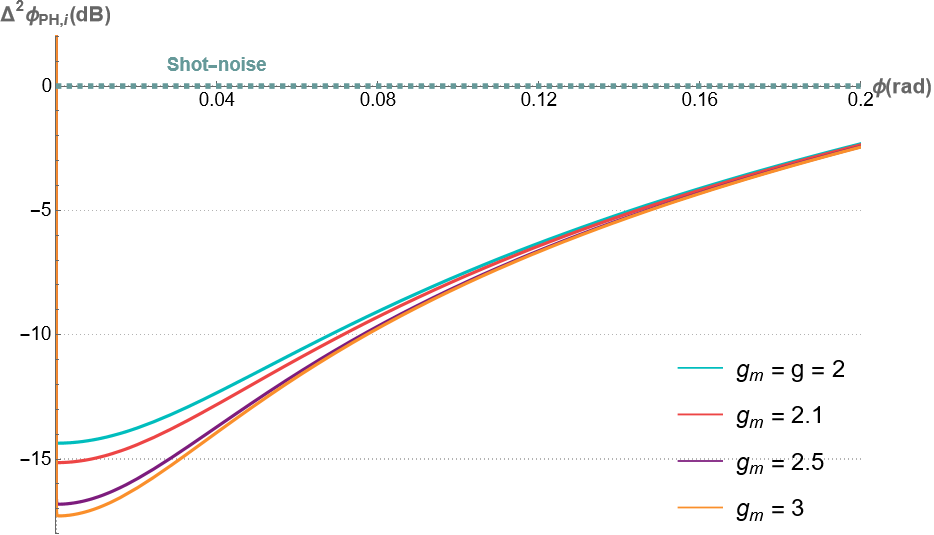}
        \caption{Single mode detection}
    \end{subfigure}%
    \begin{subfigure}{.5\textwidth}
        \centering
        \includegraphics[width=1\linewidth]{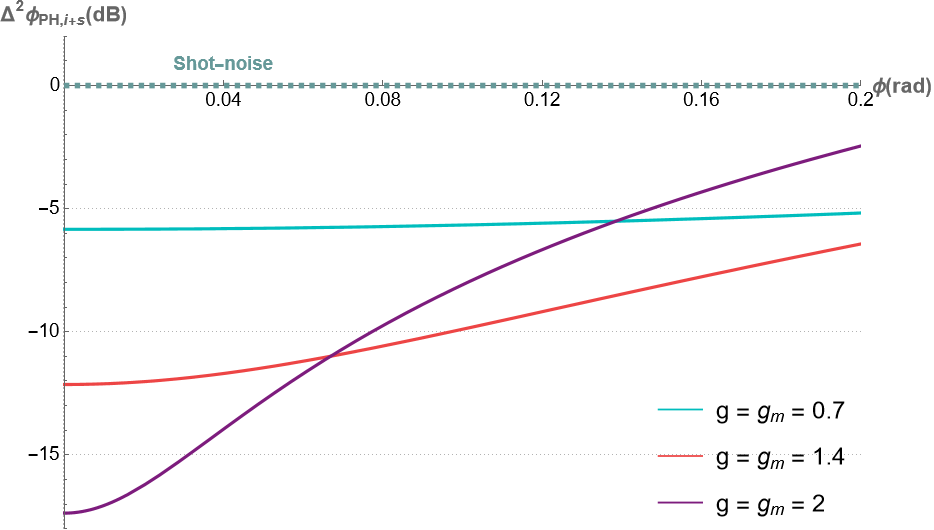}
        \caption{Dual mode detection}
    \end{subfigure}
    \caption{Calculated phase sensitivity ($\Delta^2\phi_{\text{PH}}$ in dB relative to the SNL as a function of the phase working-point $\phi$) for the non-degenerate squeezing-enhanced Sagnac interferometer with \textit{parametric homodyne detection}. We assume an ideal, lossless configuration with a coherent seed of $N_{seed}=10$. \textbf{(a) Single Mode Detection} (signal or idler) for varying measurement gain $g_m$ (fixed loop gain $g=2$) shows an enhancement of $2e^{-2g}$ for the balanced case ($g_m=g$), reaching the maximal $e^{-2g}$ enhancement when $g_m \approx 1.5g$. \textbf{(b) Dual Mode Detection} $\langle \hat{f}_{s,m}^\dagger \hat{f}_{s,m} + \hat{f}_{i,m}^\dagger \hat{f}_{i,m} \rangle$ for varying generation gain $g=0.7,1.4,2$ and balanced gain ($g_m=g$), showing the full $e^{-2g}$ enhancement.}
    \label{fig:Sensitivity_PH_detection}
\end{figure}
In parametric homodyne detection, the dark-port outputs $\hat{f}_{s/i}$ are first amplified in a second OPA (measurement OPA), whose pump phase is set to $\phi_p = \pi$ to amplify the quadrature that was initially squeezed in the loop OPA. The resulting fields are: $  \hat{f}_{s/i,m} = \hat{f}_{s/i}\cosh g_m - \hat{f}_{i/s}^{\dagger}\sinh g_m$, where $g_m$ is the gain of the measurement OPA. 

The sensitivity of non-degenerate configuration may differ from that of the degenerate due to two aspects: The measured quantity  (intensity of the signal, idler, or both) and the gain balance between the loop and the measurement OPAs: When measuring only \textit{one output mode} (signal or idler) with balanced gains ($g_m=g$) and a seed of $N_{seeed}\geq 10$, the sensitivity saturates at $2e^{-2g}$ (Fig.~\ref{fig:Sensitivity_PH_detection}a), which is 3dB less than the maximum degenerate case. However, further increasing the measurement gain to $g_m \geq 1.5g$ retrieves the maximal $e^{-2g}$ enhancement. Alternatively, when \textit{both output modes} (signal + idler) are detected simultaneously, the full $e^{-2g}$ enhancement is obtained already when the gains are balanced ($g_m = g$), as shown in Fig.~\ref{fig:Sensitivity_PH_detection}b, and further increase of the measurement gain $g_m>g$ leaves it unchanged.

Thus, while the maximum enhancement in the non-degenerate case is the same as the degenerate one, the non-degenerate scheme can enjoy the simplification of detecting only a single-mode at the cost of a higher measurement gain ($g_m \geq 1.5g$) to fully realize the $e^{-2g}$ benefit.